\documentclass[aps,prl,preprint,amsmath,amssymb]{revtex4-1}

\usepackage{graphicx}

\def\bc{\begin{center}}
\def\ec{\end{center}}
\def\be{\begin{eqnarray}}
\def\ee{\end{eqnarray}}
\newcommand{\D}{\displaystyle}

\begin{document}


\title{CONSTRAINTS FROM TYPE IA SUPERNOVAE ON \\ $\Lambda$-CDM MODEL IN RANDERS-FINSLER SPACE}



\author{Zhe Chang$^{1,2}$}
\email{zchang@ihep.ac.cn}
\author{Ming-Hua Li$^{1,2}$}
\email{limh@ihep.ac.cn}
\affiliation{${}^1$Institute of High Energy Physics, Chinese Academy
of Sciences, 100049 Beijing, China
}
\author{Xin Li$^{2,3}$}
\email{lixin@itp.ac.cn}
\affiliation{
${}^2$Theoretical Physics Center for Science Facilities, Chinese Academy of Sciences\\
${}^3$Institute of Theoretical Physics, Chinese Academy of Sciences, 100190 Beijing, China}


\date{\today}

\begin{abstract}
  Gravitational field equations in Randers-Finsler space of approximate Berwald type are investigated. A modified Friedmann equation and a new luminosity distance-redshift relation is proposed. A best-fit to the Type Ia supernovae (SNe) observations yields that the $\Omega_{\Lambda}$ in the $\Lambda$-CDM model is suppressed to almost zero. This fact indicates that the astronomical observations on the Type Ia SNe can be described well without invoking any form of dark energy. The best-fit age of the universe is given. It is in agreement with the age of our galaxy.
\end{abstract}


\maketitle

\section{I. Introdution}

The phenomenon that our universe is expanding was first observed by E. Hubble \cite{Hubble} in 1929, soon after the birth of General Relativity. Since 1990s, rapid progress on the observations of the Type Ia SNe have been made  \cite{riess}\cite{Perlmutter}\cite{Bennett}\cite{Kowalski}. In order to explain the observations, dark energy with the property of negative pressure must be involved in the framework of standard cosmological model. Multiple kinds of models have been proposed in the past decade \cite{Copeland}. Some of them suggested an evolving canonical scalar field with a potential \cite{peebles}\cite{earlyqu}\cite{Carroll:1998zi}. Others tried to modify the General Relativity to make up for the gap between the theory and the observations \cite{Capo}\cite{Carroll:2003wy}\cite{odintsovl}.  One of the most famous candidates of dark energy is the cosmological constant $\Lambda$. However, it requires ``fine tuning'' in an early epoch of the universe \cite{Steinhardt}.

\emph{Finsler geometry}, which includes the Riemann geometry as its special case, supplies a new systematic approach to the problems mentioned above. Gravity in a Finsler space has been studied for a long time \cite{Takano}\cite{Tavakol}\cite{Bogoslovsky1}. A theory of the gauge transformations in the context of Finsler space has been discussed by G. Asanov \cite{Asanov1}\cite{Asanov2}\cite {Asanov3} and S. Ikeda \cite{Ikeda1,Ikeda2}, and its applications to General Relativity has been suggested by R. Beil \cite{Beil1,Beil2}. The corresponding gravitational field equation was derived from the Riemannian osculating metric in \cite {Asanov1}. Considering consistency with the Bianchi identity in Finsler geometry and the general covariance principle of Einstein, we gave a new gravitational field equation in a Berwald-Finsler space \cite{Lixin}. In the framework of Finsler geometry, a modified Newton¡¯s gravity has been proposed, which agrees quite well with the asymptotically flat rotation curves of spiral galaxies without invoking dark matter \cite{Finsler DM}.

In this paper, we test possible constraints from the Type Ia SNe on cosmology in Finsler geometry. Based on preceding work \cite{Lixin,Modified Friedmann model}, we present a new relation between the luminosity distance and redshift.
A best-fit to the Type Ia SNe data indicates that our universe is matter-dominated in Finsler space. A reasonable age of our universe is proposed, which is consistent with that obtained from radioactive dating of isotopes in stars and from white dwarfs in our galaxy.

The paper is organized as follows. In section II, a brief introduction of Finsler geometry is presented. In section III, we discuss the gravitational field equation in a Randers-Finsler space of approximate Berwald type. A new luminosity-redshift relation is set up in section IV. It is one of the keys to understand the Type Ia SNe observations without invoking dark energy hypothesis. The numerical study is carried out in section V. In section VI, we give concluding remarks and the comparison bewtween our age prediction and that of other models.


\section{II. Finsler Geometry}
Let us first introduce basic concepts and notions in Finsler geometry. We use $T_xM$ to denote the \emph{tangent space} at a point $x$ on a manifold $M$, i.e. $x\in M$. $TM$ is the \emph{tangent bundle} of $M$. Each element of $TM$ is described by $(x, y)$, where $x\in M$ and $y\in T_xM$. The \emph{natural projection} $\pi : TM\rightarrow M$ is defined as $\pi(x, y)\equiv x$.

\emph{Finsler geometry} has its genesis in integrals of the form
\begin{eqnarray}
 \int^b_a F\left(x^1,\cdots,x^n; y^1,\cdots,y^n\right)dt\ ,
 \label{integral length}
 \end{eqnarray}
 where $x^{i}$ stands for \emph{position} and $y^i \equiv dx^{i}/{dt}$ for \emph{velocity}. The integrand $F$ is called a \emph{Finsler structure}.

A \emph{Finsler structure} of $M$ \cite{Book by Bao}
\be F :
TM\rightarrow[0,\infty)\nonumber \ee
is a function with the following properties:\\
(i) \emph{Regularity}: F is $C^\infty$ on the entire \emph{slit tangent bundle}
$TM\backslash0$.\\
(ii) \emph{Positive homogeneity} : $F(x, \lambda y)=\lambda F(x,
y)$ for all $\lambda>0$.\\
(iii) \emph{Strong convexity}: The $n\times n$ Hessian matrix
\be
g_{\mu\nu}\equiv\frac{\partial}{\partial
y^\mu}\frac{\partial}{\partial
y^\nu}\left(\frac{1}{2}F^2\right)
\label{g}
\ee is positive-definite at every point of $TM\backslash0$. And the lowering and raising of indices in this paper are carried out by the \emph{fundamental tensor} $g_{\mu\nu}$ defined above and its inverse $g^{\mu\nu}$. The \emph{Carten tensor} $A_{\lambda\mu\nu}$, which is defined as
\be
A_{\lambda\mu\nu}\equiv\frac{F}{4}\frac{\partial}{\partial y^\lambda}\frac{\partial}{\partial y^\mu}\frac{\partial}{\partial y^\nu}(F^2)\ ,
\label{carten}
\ee
is regarded as a measurement of deviation from the Riemannian manifold.

According to Chern's work \cite{Chern2}, each Finsler manifold admits a unique linear connection, called the \emph{Chern connection}. It is torsion-free and almost metric-compatible. The \emph{connection coefficients} take a form as
 \be
 \Gamma^{\alpha}_{~\mu\nu}=\gamma^{\alpha}_{~\mu\nu}-g^{\alpha\lambda}\left(A_{\lambda\mu\beta}\frac{N^\beta_{~\nu}}{F}-A_{\mu\nu\beta}\frac{N^\beta_{~\lambda}}{F}+A_{\nu\lambda\beta}\frac{N^\beta_{~\mu}}{F}\right),
 \label{Gamma}
  \ee
 where $\gamma^{\alpha}_{~\mu\nu}$ is the \emph{formal Christoffel symbols} of the second kind
 \be
 \gamma^{\alpha}_{~\mu\nu} \equiv  \frac{1}{2}g^{\alpha s} \left(\frac{\partial g_{s\mu}}{\partial x^{\nu}} - \frac{\partial g_{\mu\nu}}{\partial x^{s}} + \frac{\partial g_{\nu s}}{\partial x^{\mu}}\right) \ .
 \label{gamma2}
 \ee
$N^\mu_{~\nu}$ is defined as
\be
N^\mu_{~\nu} \equiv \gamma^\mu_{~\nu\alpha}y^\alpha- \frac{A^\mu_{~\nu\lambda}}{F}\gamma^\lambda_{~\alpha\beta}y^\alpha
y^\beta \ .
\label{N}
\ee

The \emph{curvature tensor} of a Finsler space is given as \cite{Book by Bao}
\be
R^{~\lambda}_{\kappa~\mu\nu}&=&\frac{\delta
\Gamma^\lambda_{~\kappa\nu}}{\delta x^\mu}-\frac{\delta
\Gamma^\lambda_{~\kappa\mu}}{\delta
x^\nu}+\Gamma^\lambda_{~\alpha\mu}\Gamma^\alpha_{~\kappa\nu}-\Gamma^\lambda_{~\alpha\nu}\Gamma^\alpha_{~\kappa\mu}\ ,
\ee
where
\be
\frac{\delta}{\delta x^\mu} = \frac{\partial}{\partial x^\mu}-N^{\nu}_{~\mu}\frac{\partial}{\partial y^\mu}\ .
\label{delta}
\ee

The \emph{Ricci scalar} is defined as
\be
Ric \equiv g^{\mu\nu}R_{\mu\nu}\ ,
\label{Ric}
\ee
where
\be
R^{\mu}_{~\nu} \equiv \frac{y^{\lambda}}{F} R^{~\mu}_{\lambda~\nu\kappa} \frac{y^{\kappa}}{F} = \frac{y^{\lambda}}{F}
\left( \frac{\delta}{\delta x^\nu} \frac{N^\mu_{~\lambda}}{F}- \frac{\delta}{\delta x^\lambda} \frac{N^\mu_{~\nu}}{F} \right)\ .
\label{Ruv}
\ee

The \emph{Ricci tensor} $Ric_{\mu\nu}$, first introduced by Akbar-Zadeh \cite{Akbar}, is
\be
Ric_{\mu\nu} \equiv \frac{\partial^2 (\frac{1}{2}F^2Ric)}{\partial y^\mu \partial y^\nu} = \left[\frac{1}{2}F^2Ric\right]_{y^\mu y^\nu}\ .
\label{Ricuv}
\ee

A \emph{Randers space} is a specific type of Finsler space, whose Finsler structure takes the form \cite{Randers}
\begin{eqnarray}
F(x,y)\equiv\alpha(x,y)+\beta(x,y)\ ,
\label{rander}
\end{eqnarray}
where
\begin{eqnarray}
\alpha(x,y)&\equiv&\sqrt{\tilde{a}_{\mu\nu}(x)y^\mu y^\nu}\ ,
\end{eqnarray}
\begin{eqnarray}
\beta(x,y)&\equiv&\tilde{b}_\mu(x)y^\mu\ .
\end{eqnarray}
The $\tilde{a}_{\mu\nu}$ are the components of a Riemannian metric and the $\tilde{b}_\mu$ are those of a 1-form. Lower case Greek indices run from `$0$' to `$3$'. Specifically, the lowering and raising of indices for the terms decorated with a tilde are carried out by $\tilde{a}_{\mu\nu}$ and its inverse $\tilde{a}^{\mu\nu}$ instead of the fundamental tensor.

A Finsler structure $F$ is said to be of \emph{Berwald type} if the Chern connection coefficients $\Gamma^{\alpha}_{~\mu\nu}$ in natural coordinates have no $y$ dependence. A Randers space is said to be of
Berwald type if and only if \cite{Kikuchi}
\be
\label{condition of RB}
\tilde{b}_{\mu|\nu}\equiv\frac{\partial\tilde{b}_\mu}{\partial
x^\nu}-\tilde{b}_\kappa\tilde{\gamma}^\kappa_{~\mu\nu}=0 \ ,
\ee
where $\tilde{\gamma}^\kappa_{~\mu\nu}$ is the Christoffel symbols of a Riemannian metric $\tilde{\alpha}$. After some tedious calculations, one obtains that
\be
\Gamma^\kappa_{~\mu\nu}=\tilde{\gamma}^\kappa_{~\mu\nu}\ .
\ee
So the curvature tensor of a Randers space of Berwald type is given as
\be R^{~\lambda}_{\kappa~\mu\nu}&=&\frac{\partial
\tilde{\gamma}^\lambda_{~\kappa\nu}}{\partial x^\mu}-\frac{\partial
\tilde{\gamma}^\lambda_{~\kappa\mu}}{\partial
x^\nu}+\tilde{\gamma}^\lambda_{~\alpha\mu}
\tilde{\gamma}^\alpha_{~\kappa\nu}-\tilde{\gamma}^\lambda_{~\alpha\nu}\tilde{\gamma}^\alpha_{~\kappa\mu}\ ,
\label{curvature}
\ee
and the corresponding Ricci tensor $Ric_{\mu\nu}$ is
\be
Ric_{\mu\nu}=\frac{1}{2}(R^{~\alpha}_{\mu~\alpha\nu}+R^{~\alpha}_{\nu~\alpha\mu})\ .
\label{Ricci}
\ee

\section{III. The Friedmann Model in the framework of Finsler Geometry}
In order to investigate the Friedmann-Robertson-Walker (FRW) cosmology, we take $\tilde{\alpha}_{\mu\nu}$ to be the form
\be
\tilde{a}_{\mu\nu}={\rm diag}\left(1, -\frac{R^2(t)}{1-kr^2}, -R^2(t)r^2, -R^2(t)r^2\sin^2\theta\right)\ ,
\ee
where $k=0,+1,-1$ stands for a flat, closed or open universe respectively. With the condition (\ref{condition of RB}) in mind and assuming that the space of our universe is almost homogeneous and isotropic, we take
\be
\tilde{b}_\mu=(\tilde{b}_0,0,0,0) \ ,
\ee
where $\tilde{b}_0$ is a small constant.

Using the identities (\ref{g}), (\ref{curvature}) and (\ref{Ricci}), one may directly calculates the Ricci tensor in the Randers space of approximate Berwald type. Nonzero components are listed below:
\be
Ric_{00}&=&-3\frac{\ddot{R}}{R}\tilde{a}_{00},\nonumber\\
Ric_{ij}&=&-\left(\frac{\ddot{R}}{R}+2\frac{\dot{R}^2}{R^2}+\frac{2k}{R^2}\right)\tilde{a}_{ij}\ .
\label{Ric nonzero}
\ee
The trace of the Ricci tensor $Ric_{\mu\nu}$ gives the \emph{scalar curvature} $S \equiv g^{\mu\nu}Ric_{\mu\nu}$ ,
\be
S&=&-6\frac{\alpha}{F}\left(\frac{\ddot{R}}{R}+\frac{\dot{R}^2}{R^2}+\frac{k}{R^2}\right) -3\frac{\ddot{R}}{R}\frac{\alpha^2}{F^2}\left(\frac{\beta}{F}\tilde{a}_{00}\frac{y^0}{\alpha}\frac{y^0}{\alpha}-2\tilde{a}_{00}\frac{y^0}{\alpha}\tilde{b}^0\right)\nonumber\\
&&-\left(\frac{\ddot{R}}{R}+2\frac{\dot{R}^2}{R^2}+\frac{2k}{R^2}\right)\frac{\alpha^2}{F^2}\left(\frac{\beta}{F}\tilde{a}_{ij}\frac{y^i}{\alpha}\frac{y^j}{\alpha}\right).
\label{S}
\ee

A new gravitational field equation in the Berwald-Finsler space is given as \cite{Lixin}
\begin{eqnarray}
\left[Ric_{\mu\nu}-\frac{1}{2}g_{\mu\nu}S\right]+\left\{\frac{1}{2}
B^{~\alpha}_{\alpha~\mu\nu}+B^{~\alpha}_{\mu~\nu\alpha}\right\}=8\pi
G T_{\mu\nu}\ ,
\label{new field equation}
\end{eqnarray}
where
\be
B_{\mu\nu\alpha\beta}=-A_{\mu\nu\lambda}R^{~\lambda}_{\theta~\alpha\beta}y^\theta/F\ .
\ee
$T^\mu_\nu$ is the energy-momenta tensor as $T^\mu_\nu={\rm diag}(\rho, -p,-p,-p)$, where $p\equiv p(x)$ and $\rho\equiv\rho(x)$ is  respectively the pressure and the energy density of the cosmic fluid.

In a Randers space of approximate Berwald type, the gravitational field equation (\ref{new field equation}) reduces to
\begin{eqnarray}
\left[Ric_{\mu\nu}-\frac{1}{2}g_{\mu\nu}S\right] = 8\pi G T_{\mu\nu}\ ,
\label{field equation of Berwald}
\end{eqnarray}
because the terms $B^{~\alpha}_{\alpha~\mu\nu}$ and $B^{~\alpha}_{\mu~\nu\alpha}$ are zero. $Ric_{\mu\nu}$ and $S$ are given by the identities (\ref{Ric nonzero}) and (\ref{S}).

For the sake of simplicity, we introduce two parameters $A$ and $B$ as
\be
A \equiv \frac{\alpha}{F}\left(\frac{\beta}{F}\tilde{a}_{00}\frac{y^0}{\alpha}\frac{y^0}{\alpha} - 2\tilde{a}_{00}\frac{y^0}{\alpha}\tilde{b}^{0}\right)
\ee
and
\be
B \equiv \frac{\alpha}{F}\left(\frac{\beta}{F}\tilde{a}_{ij}\frac{y^i}{\alpha}\frac{y^j}{\alpha}\right)\ .
\ee
The $0$-$0$ component of the field equation (\ref{field equation of Berwald}) gives the modified Friedmann equation
\begin{eqnarray}
\frac{\alpha}{F}\left(\frac{\dot{R}^2}{R^2}+\frac{k}{R^2}\right)-\frac{1}{2}\frac{\alpha}{F}\frac{\ddot{R}}{R}A
+ \frac{1}{6}\frac{\alpha}{F}\left(\frac{\ddot{R}}{R}+2\frac{\dot{R}^2}{R^2}+2\frac{k}{R^2}\right)B=\frac{8\pi
G}{3}\rho
\label{0-0}
\end{eqnarray}
and the $i$-$j$ component of (\ref{field equation of Berwald}) gives
\begin{eqnarray}
3\frac{\alpha}{F}\left(2\frac{\ddot{R}}{R}+\frac{\dot{R}}{R^2}+\frac{k}{R^2}\right) + \frac{9}{2}\frac{\ddot{R}}{R}\frac{\alpha}{F}A + \frac{1}{2}\frac{\alpha}{F}\left(\frac{\ddot{R}}{R}+2\frac{\dot{R}}{R^2}+2\frac{k}{R^2}\right)B = 8\pi
G(-3p)\ .
\label{i-i}
\end{eqnarray}

Subtracting the above two equations, we have
\be
\frac{\alpha}{F}\frac{\ddot{R}}{R}(1+A)=-\frac{4\pi G}{3}(\rho+3p)\ .
\label{1}
\ee
The $0$-$0$ component of the field equation (\ref{0-0}) can be rewritten into the form
\be
\frac{\alpha}{F}\left(\frac{\dot{R}^2}{R^2}+\frac{k}{R^2}\right)\left(1+\frac{B}{3}\right)\left(1+A\right)=\frac{8\pi G}{3}\rho\left(1+A\right)+ \frac{4\pi G}{3}\left(\rho+3p\right)\left(-\frac{A}{2}+\frac{B}{6}\right)\ .
\label{key}
\ee

Assuming that $A$ and $B$ are both time-independent, implementation of time derivative $\frac{d}{dt}$ on both sides of the equation (\ref{key}) and using (\ref{1}) again leads us to
\be
-\frac{\dot{R}}{R}\left[\left(1+\frac{B}{3}\right)(\rho+3p)+\rho\left(2+\frac{3A}{2}+\frac{B}{6}\right)+p\left(-\frac{3A}{2}+\frac{B}{2}\right) \right]=\  &&\dot{\rho}\left(1+\frac{3A}{4}+\frac{B}{12}\right)\nonumber\\
&&+\dot{p}\left(-\frac{3A}{4}+\frac{B}{4}\right)\ .
\label{4}
\ee
With the equations of state $p_i = w_i \rho_i$ of each individual component $i$ (where the constant $w_i=0,-1,-1/3$ corresponds to \emph{matter}, \emph{vacuum} and `\emph{curvature}' respectively), the equation (\ref{4}) can be solved,
\be
\rho_i \propto R^{-\D{3(1+w_i)+\frac{3}{2}(1-w_i)A+\frac{1}{2}(1+3w_i)B \over 1+\frac{3}{4}(1-w_i)A+\frac{1}{12}(1+3w_i)B}}\ .
\label{rho}
\ee

\section{IV. A New Luminosity-Redshift Relation}
We adopt the conventional definitions
\begin{eqnarray}
H(a) \equiv  \frac{\dot{R}}{R}\ , \qquad
\rho_{\rm crit0} \equiv {{3H_0^2}\over{8\pi G}}\ , \qquad
\Omega_{i0} \equiv {{\rho_{i0}}\over {\rho_{\rm crit0}}}=\left({{8\pi G}\over 3H_0^2}\right) \rho_{i0}\ ,
\end{eqnarray}
and
\begin{eqnarray}
  \rho_k \equiv -{{3k}\over{8\pi GR_0^2 a^{-2}}}\ ,\qquad
  \rho_{\rm vac} = \rho_\Lambda \equiv {{\Lambda}\over{8\pi G}}\ .
\end{eqnarray}
Combining the equations (\ref{key}) and (\ref{rho}), one obtains
\begin{eqnarray}
  H(a) =
  H_0  \left[ \sum_{i (k)} \Omega_{i0} f_i(w_i,A,B) a^{-n_i(w_i,A,B)} \right] ^{1/2}\ ,
\label{H(a)}
\end{eqnarray}
where
\begin{eqnarray}
   f_i(w_i,A,B) = \frac{1 + \frac{3}{4} (1-w_i)A + \frac{1}{12}(1+3w_i)B }{(1+\frac{B}{3})(1+A)}\ ,
\end{eqnarray}
\begin{eqnarray}
   n_i(w_i,A,B) = - \frac{ 3(1+w_i) + \frac{3}{2}(1-w_i)A + \frac{1}{2}(1+3w_i)B }{
   1 + \frac{3}{4}(1-w_i)A + \frac{1}{12}(1+3w_i)B }\ .
\end{eqnarray}
 Here $\Omega_{\rm k} = 1 - \Omega_{\rm M} - \Omega_{\Lambda}$ . The notation $\sum_{i (k)}$ denotes that the sum includes $\Omega_k$. As an approximation, we do not take the radiation term into account due to its little influence on predictions of the $\Lambda$-CDM model when using the Type Ia SNe data alone.

The \emph{luminosity distance} $d_L$ as a function of the redshift $z$ of a supernova is \cite{Sean M. Carroll}
\begin{eqnarray}
  \begin{array}{rcl}
  d_L(z)
   = & \displaystyle{
  {(1+z)\over{H_0\sqrt{|\Omega_{k0}|}}} \, {\rm sinn}\left[
  H_0\sqrt{|\Omega_{k0}|}\int^{1}_{1/(1+z)}
  {{da}\over{a^2 H(a)}} \right]}\ ,
  \end{array}
\label{d_L(z)}
\end{eqnarray}
where `$sinn$' stands for `$sin$' (if $k>0$), `$1$' (if $k=0$) or `$sinh$' ( if $k<0$ ) under certain circumstances.
Substituting the equation (\ref{H(a)}) into (\ref{d_L(z)}), we get a luminosity distance-redshift relation that looks like
\begin{eqnarray}
  d_L(z) = d_L(z; \Omega_{\rm M},\Omega_\Lambda, A, B)  \ .
\label{d_L(z;A,B)}
\end{eqnarray}


\section{V. Numerical Study}
The \emph{distance modulus} $\mu$ is related to the luminosity distance via
\begin{eqnarray}
    \mu \equiv m - M = 5 \log_{10}[d_L({\rm Mpc})] + 25\ ,
\label{distance modulus}
\end{eqnarray}
where $m$ is the \emph{apparent magnitude} of the source and $M$
its \emph{absolute magnitude}. And it is $\mu$ and $z$ that the Supernova Project measured. A total uncertainty of $\mu$ (denoted by `$\sigma_\mu$') and the corresponding redshift $z$ were presented in the reference \cite{Kowalski}.

The $\chi^2$ statistic in our fit is
\begin{eqnarray}
\chi_{\rm SN}^2(\Omega_{\rm M},\Omega_{\Lambda}) \equiv \sum_{i=1}^{307}{[\mu_{\rm obs}(z_i)-
\mu_{\rm th}(z_i;\Omega_{\rm M},\Omega_{\Lambda})]^2 \over {{\sigma_{\mu}(z_i)}^2}}\ ,
\end{eqnarray}
where $\mu_{\rm th}(z_i)$ is obtained by the equation(\ref{distance modulus}), while $\mu_{\rm obs}(z_i)$ and $\sigma_{\mu}(z_i)$ come from the observations. We employ the Markov Chain Monte-Carlo (MCMC) techniques \cite{Antony Lewis} to explore the parameter space. The likelihood function looks like ${\cal L} \propto e^{-\chi_{\rm SN}^2(\Omega_{\rm M},\Omega_{\Lambda})/2}$. For simplicity, we take $A= -3$ and $B= -1$,
leaving $\Omega_\Lambda$(or $\Omega_{\rm M}$) to be the \emph{only free parameter} in our model with the constraint $\Omega_{\rm M} + \Omega_\Lambda = 1$ of a $k = 0$ flat universe.



\begin{figure}
\begin{center}
\scalebox{0.62}[0.57]{\includegraphics{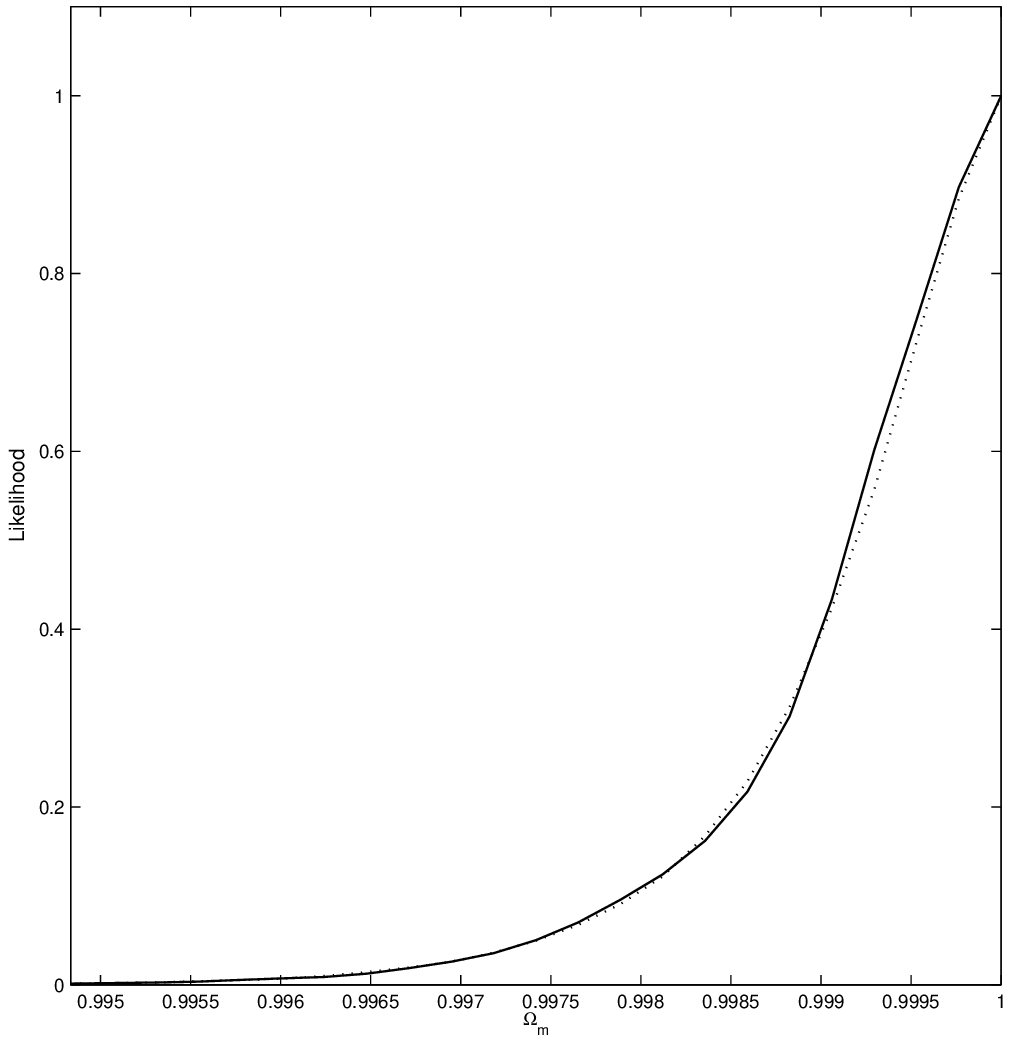}}
\caption{The posterior constraints of $\Omega_{\rm M}$ with $A = -3$ and $B = -1$ using all data. The solid line indicates the fully marginalized posterior of $\Omega_{\rm M}$. The dotted line shows the mean likelihood of the samples. The fact that the two matching well implies that the mean likelihood is well constrained and our result is justifiable. We take the Hubble constant to be $H_0 = 70.5$ $\rm km\cdot s^{-1}\cdot Mpc^{-1}$ instead of a base variable parameter, because the Type Ia supernova data alone cannot put a well constraint on it. The center value lies at $\Omega_{\rm M}= 0.9997$. The best-fit likelihood ${\cal L}$ for a total number of $307$ data points is $-ln{\cal L} ({\cal L} \propto e^{-\chi_{\rm SN}^2(\Omega_{\rm M},\Omega_{\Lambda})/2}) =182.1819$. Corresponding to a high best-fit $\Omega_{\rm M}$, a low best-fit $\Omega_{\Lambda} = 0.0003$ should be anticipated, which indicates a matter-dominated universe.}
\label{fig-3}
\end{center}
\end{figure}

\begin{figure}
\begin{center}
\scalebox{0.65}[0.53]{\includegraphics{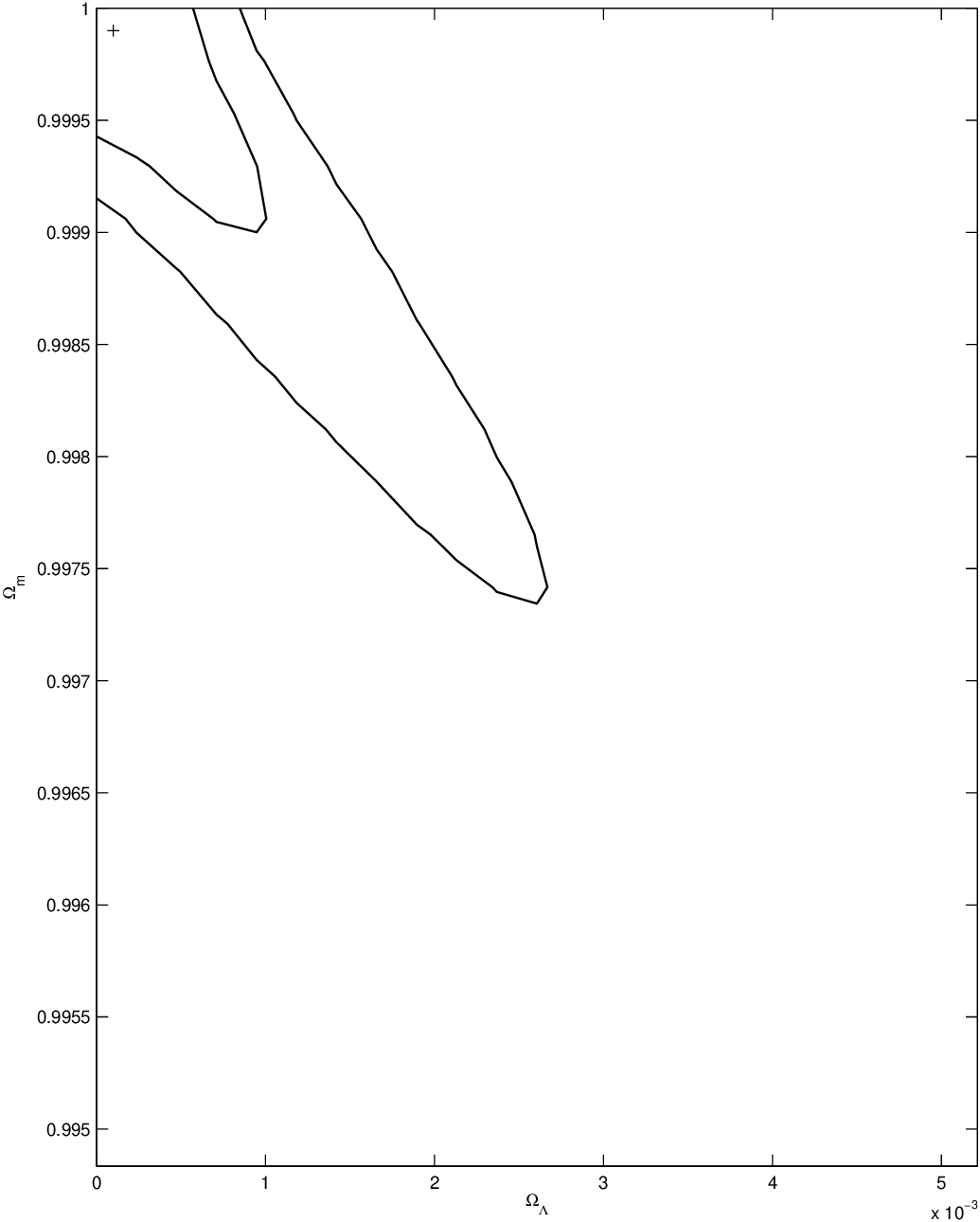}}
\caption{Best-fit $68\%$ and $95\%$ confidence regions from the marginalized posterior distributions in the $\Omega_\Lambda$-$\Omega_{\rm M}$ plane for the Finsler cosmological model with $A = -3$ and $B = -1$. The inner contour denotes the $68\%$ confidence limit and the outer one denotes the $95\%$ one. The cross `$+$' at the upper left corner denotes the best-fit values of $(\Omega_\Lambda, \Omega_{\rm M})=(0.0003,0.9997)$ in the modified Friedmann model. Compared to the $(\Omega_{\Lambda}, \Omega_{\rm M})=(0.28,0.72)$ in the reference \cite{spergel}, our result indicates that in a Randers-Finsler universe, no dark energy but only matter components exist.}
\label{fig-4}
\end{center}
\end{figure}

\begin{figure}
\begin{center}
\scalebox{0.65}[0.53]{\includegraphics{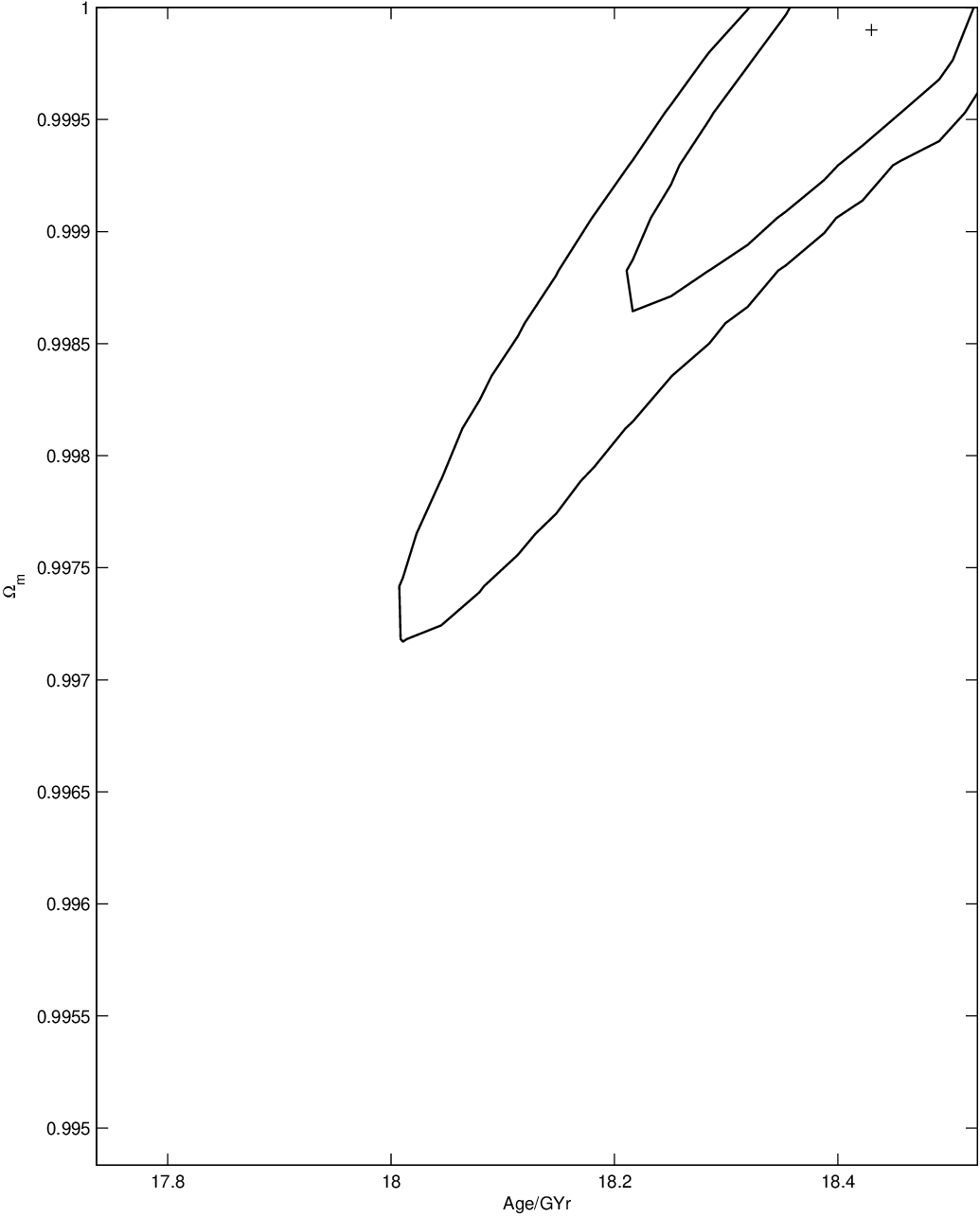}}
\caption{Best-fit $68\%$ and $95\%$ confidence regions from the marginalized posterior distributions in the Age-$\Omega_{\rm M}$ plane for the Finsler cosmological model with $A = -3$ and $B = -1$. The inner contour denotes the $68\%$ confidence limit and the outer one denotes the $95\%$ one. The cross `$+$' at the upper right corner in the above figure denotes the best-fit value of the age of the universe in our modified Friedmann model is $18.298$ Gyr. This prediction is consistent with the $10$ to $20$ Gyr estimate obtained from radioactive dating of isotopes in stars \cite{schramm}\cite{truran} and the $6.5$ to $10$ Gyr minimum age given by the white dwarfs in our Galactic disk \cite{oswalt96}\cite{bergeron}. Thus, the Type Ia SNe data could be well explained by a Randers-Finsler space-time without invoking any form of dark energy.}
\label{fig-2}
\end{center}
\end{figure}

The marginalized posterior and the mean likelihood distributions of the density parameter $\Omega_{\rm M}$ are shown in Fig.[\ref{fig-3}]. The two contours in Fig.[\ref{fig-4}] and Fig.[\ref{fig-2}] respectively line out the $68\%$ and $95\%$ confidence regions of the marginalized distribution in the $\Omega_\Lambda$-$\Omega_{\rm M}$ and Age-$\Omega_{\rm M}$ planes. Best-fit values of $\Omega_{\rm M} = 0.9997_{-0.0009}^{+0.0003}$ and $\Omega_\Lambda = 0.0003_{-0.0003}^{+0.0001}$ are obtained with $-ln{\cal L} = 182.1819$ for a total number of $307$ data points. The almost vanished $\Omega_\Lambda$ indicates that, in our model, there is no need to invoke the $\Omega_\Lambda$ term in the Einstein's field equation to account for the supernova observations. The best-fit age of the universe is $18.298_{-0.013}^{+0.102}$ Gyr.


\section{VI. Conclusions}
In this paper, we have initiated an exploration on the possibility of a modified Friedmann model in a Randers-Finsler space of approximate Berwald type as an alternative to the dark energy hypothesis. Wondering whether the space-time of our universe is a Randers-Finslerian manifold instead of a Riemannian one, we have rewritten the Einstein's field equation in such a space and the new Friedmann equation was also given. A best-fit to the Type Ia SNe data suppresses the effective density parameter $\Omega_{\Lambda}$ in the $\Lambda$-CDM model to almost zero. This fact demonstrates that a Randers-Finsler geometrical explanation of the `accelerated' expanding universe without invoking dark energy is possible.


Moreover, the best-fit age of the universe in our model is consistent with the 10 to 20 Gyr estimate obtained from radioactive dating of isotopes in stars \cite{schramm}\cite{truran} and the 6.5 to 10 Gyr minimum age given by the white dwarfs in our Galactic disk \cite{oswalt96}\cite{bergeron}. However, the change from the old Riemann space-time to a new Randers-Finsler one may call for a redefinition of not only the luminosity distance, but also
probably other metric-related quantities. This will be the subject of our future investigation.
\begin{acknowledgments}
\section{Acknowledgments}
Our work was supported by the NSF of China under Grant No. 10575106 and No. 10875129.

\end{acknowledgments}

\bibliography{basename of .bib file}

\begin{thebibliography}{999}
\bibitem{Hubble}E. Hubble, {\it Proceedings of the National Academy of Sciences of the United States of America} {\bf 15}, 168 (1929).
\bibitem{riess}
A.~Riess {\it et al.},
Astron.\ J.\  {\bf 116}, 1009 (1998);
Astron.\ J.\  {\bf 117}, 707 (1999).
\bibitem{Perlmutter}S. Perlmutter {\it et al}., Astrophys J. {\bf 517}, 565 (1999).
\bibitem{Bennett}C. Bennett {\it et al}., Astrophys J. Suppl. {\bf 148}, 1 (2003).
\bibitem{Kowalski}M. Kowalski {\it et al}., Astrophys. J. {\bf 686}, 749 (2008).

\bibitem{Copeland}E. Copeland {\it et al}., Int. J. Mod. Phys. D {\bf 15}, 1753 (2006).

\bibitem{peebles}
B.~Ratra and J.~Peebles, Phys. \ Rev. \ D {\bf 37},
321 (1988).

\bibitem{earlyqu}
Y.~Fujii,
Phys.\ Rev.\ D {\bf 26}, 2580 (1982);
L.~H.~Ford,
Phys.\ Rev.\ D {\bf 35}, 2339 (1987);
Y.~Fujii and T.~Nishioka,
Phys.\ Rev.\ D {\bf 42}, 361 (1990);
T.~Chiba, N.~Sugiyama and T.~Nakamura,
Mon.\ Not.\ Roy.\ Astron.\ Soc.\  {\bf 289}, L5 (1997).

\bibitem{Carroll:1998zi}
S. ~Carroll,
Phys.\ Rev.\ Lett.\  {\bf 81}, 3067 (1998).

\bibitem{Capo}
S.~Capozziello, S.~Carloni and A.~Troisi,
arXiv:astro-ph/0303041;
S.~Capozziello, V.~Cardone, S.~Carloni, and A.~Troisi,
Int.\ J.\ Mod.\ Phys.\ D {\bf 12}, 1969 (2003).

\bibitem{Carroll:2003wy}
S. Carroll, V. Duvvuri, M. Trodden, and M. Turner,
Phys.\ Rev.\ D {\bf70}, 043528  (2004).

\bibitem{odintsovl}
S.~Nojiri and S.~D.~Odintsov,
arXiv:hep-th/0601213.
\bibitem{Steinhardt}P. Steinhardt, {\it Critical Problems in Physics}, edited by V. L. Fitch and D. R. Marlow, Princeton University Press, Princeton, NJ, 1997.
\bibitem{Takano}Y. Takano, Lett. Nuovo Cimento {\bf 10}, 747
(1974).
\bibitem{Tavakol}R. K. Tavakol, N. Van Den Bergh, Phys. Lett. A {\bf 112}, 23
(1985).
\bibitem{Bogoslovsky1}G. Yu. Bogoslovsky, Phys. Part. Nucl. {\bf 24}, 354
(1993).
\bibitem{Asanov1}G. Asanov, {\it Finsler Geometry, Relativity and Gauge Theories}, Reidel
Pub.Com., Dordrecht, 1985.
\bibitem{Asanov2}G. Asanov, Annalen der Physik {\bf 44}, 1 (1987).
\bibitem{Asanov3}G. Asanov and M. Kiselev, Rep. Math. Phys., {\bf 26}, 401 (1988).
\bibitem{Ikeda1}S. Ikeda, J. Math. Phys. {\bf 26}, 958 (1985).
\bibitem{Ikeda2}S. Ikeda, Annalen der Physik {\bf 46}, 173 (1989).
\bibitem{Beil1}R. Beil, Int. J. Theor. Phys. {\bf 30}, 1663 (1991).
\bibitem{Beil2}R. Beil, Int. J. Theor. Phys. {\bf 31}, 1025 (1992).
\bibitem{Lixin}X. Li and Z. Chang, Chinese Phys. C {\bf 34}, 28 (2010).
\bibitem{Finsler DM}Z. Chang and X. Li, Phys. Lett. B {\bf 668}, 453 (2008).
\bibitem{Modified Friedmann model}Z. Chang and X. Li, Phys. Lett. B {\bf 676}, 173 (2009).
\bibitem{Book by Bao}D. Bao, S. S. Chern and Z. Shen, {\it An Introduction to Riemann--Finsler Geometry}, Graduate Texts in Mathmatics {\bf 200}, Springer, New York, 2000.
\bibitem{Chern}S. S. Chern, {\it Finsler geometry is just Riemann geometry without the quadratic restrictions}, Notice of AMS, 959 (1996).
\bibitem{Chern2}S. S. Chern, Sci. Rep. Nat. Tsing Hua Univ. Ser. A {\bf 5}, 95 (1948); or Selected Papers, vol. II, 194, Springer, 1989.
\bibitem{Randers}G. Randers, Phys. Rev. {\bf 59}, 195 (1941).

\bibitem{Kikuchi}S. Kikuchi, Tensor, N.S. {\bf 33}, 242 (1979).
\bibitem{Akbar}H. Akbar-Zadeh, Acad. Roy. Belg. Bull. Cl. Sci. (5) {\bf 74}, 281 (1988).
\bibitem{Sean M. Carroll}S. Carroll, Living Rev. Relativity {\bf 4}, 1 (2001).
\bibitem{Antony Lewis}A. Lewis, Phys. Rev. D {\bf 66}, 103511 (2002).
\bibitem{schramm}D. Schramm, in {\em Astrophysical Ages and Dating Methods},
edited by E. Vangioni-Flam {\it et al.}, Gif sur Yvette: Edition Frontieres: Paris, 1989.
\bibitem{truran} J. Truran, in {\it The Extragalactic Distance Scale},
edited by M. Livio, M. Donahue and N. Panagia, Cambridge University Press, 1997,
p. 18.
\bibitem{oswalt96} T. Oswalt, J. Smith, M. Wood \& P. Hintzen, Nature {\bf 382}, 692 (1996).
\bibitem{bergeron} J. Bergeron {\it et al}., Astrophys. J. Suppl. {\bf 108}, 339 (1997).
\bibitem{spergel}D. Spergel {\it et al}., Astrophys. J. Suppl. {\bf 170}, 377 (2007).


\end{thebibliography}

\end{document}